\documentclass[prb,twocolumn,groupeaddress]{revtex4-2}

\usepackage{amsmath}
\usepackage{graphicx}
\usepackage[colorlinks=true, allcolors=blue]{hyperref}
\usepackage{bbold}

\begin{document}

\title{Manifest charge-transfer physics in T$^\prime$-La$_2$NiO$_4$}
\author{Wiktoria Lewandowski}
\affiliation{Theoretische Physik III, Ruhr-Universit\"at Bochum,
  D-44780 Bochum, Germany}
\author{Steffen B\"otzel}
  \affiliation{Theoretische Physik III, Ruhr-Universit\"at Bochum,
  D-44780 Bochum, Germany}
\author{Ilya M. Eremin}
\affiliation{Theoretische Physik III, Ruhr-Universit\"at Bochum,
  D-44780 Bochum, Germany}
\author{Frank Lechermann}
\affiliation{Theoretische Physik III, Ruhr-Universit\"at Bochum,
  D-44780 Bochum, Germany}

\begin{abstract}
  The La$_2$NiO$_4$ compound stabilizes in nature in the so-called T structure with
  octahedral oxygen coordination of the Ni site. Motivated by a similarly
  existing polymorph for La$_2$CuO$_4$ in the cuprate system, we here study La$_2$NiO$_4$ in the
  T$^\prime$-structure with square-planar oxygen coordination of Ni by means
  of first-principles many-body theory. The hypothetical T$^\prime$-La$_2$NiO$_4$ compound turns out
  to be a manifest charge-transfer insulator with a cuprate-like charge gap of $\sim 1.8$\,eV.
  Upon doping, a strong carrier asymmetry is revealed, i.e. while holes dominantly enter O$(2p)$-derived
  states, electrons
  may become itinerant majorly within effective Ni-$d_{x^2-y^2}$ states. Nearest-neighbor antiferromagnetic
  ordering at stoichiometry appears absent in the T$^\prime$-structure, in stark contrast to the
  strong antiferromagnetism in the T structure. Our theoretical study opens up new pathways for
  correlation physics in layered nickelates with challenging charge-transfer signatures, 
  awaiting experimental inquiries.
\end{abstract}

\maketitle
\textit{Introduction.---}
The single-layer compound La$_2$NiO$_4$ from the associated Ruddlesden-Popper (RP) nickelate series
is one of the founding fathers of the materials research on strongly correlated electron
systems~\cite{goo73}. It is isostructural with the canonical high-$T_c$ cuprate La$_2$CuO$_4$,
and has been subject of numerous experimental and theoretical studies (see e.g.
Ref.~\cite{ima98} for a review). Generally, it has long been known~\cite{tok89,cho90,mat09}
that such single-layer RP oxides may come in two different polymorphs.
Namely the (usually stable) T structure with octahedral oxygen coordination of the
transition-metal site, and a (usually metastable) T$^\prime$ structure with square-planar
oxygen coordination of the transition-metal site. Standard T-La$_2$CuO$_4$ is a
charge-transfer insulator, whereas the corresponding T$^\prime$ structure shows finite
conductivity. While nominally of perfect stoichiometry, T$^\prime$-La$_2$CuO$_4$ gives rise to
signatures of electron doping in its electronic structure, possibly due to intrinsic defect
formation~\cite{bri95,sek03,wei16}. In spite of these interesting physics
relations from the cuprate system, to our knowledge no reports of T$^\prime$-La$_2$NiO$_4$ exist
in the literature. However, also from the vantage point of recent superconductivity in layered
nickelates~\cite{li19,sun23}, scrutinizing this single-layer polymorph may be worthwhile.

Bulk-stable T-La$_2$NiO$_4$ crystallizes in an orthorhombic lattice structure. It
is a nearest-neighbor (NN) antiferromagnetic (AFM) Mott/charge-transfer insulator
with a sizeable charge gap
of about 4\,eV and an ordered moment of $\sim 1.7\,\mu_{\rm B}$~\cite{lan89,rod91,eis92}.
The correlated electronic structure of this nominal Ni$(3d^8)$ compound is dominated by half-filled
Ni-$e_g$ $\{d_{x^2-y^2},d_{z^2}\}$ orbitals~\cite{guo88,par12,lan20,lec22,lab24,tang25},
with a charge-transfer energy $\Delta=\varepsilon_d-\varepsilon_p=3.9$\,eV. While already a small
amount of hole doping destroys the AFM order, (bad) metallicity is only reached for
substantial doping levels~\cite{tak90,sre90,cav91,shi02,ban23}. Recently, the T$^\prime$ compound
La$_2$NiO$_3$F has been synthesized by topochemical defluorination of La$_2$NiO$_3$F$_2$~\cite{wis20}.
A density-functional theory (DFT) study of this compound reveals a slightly reduced
$\Delta=3.6$\,eV and effective Ni-$d_{x^2-y^2}$ single-band metallicity from the weak-coupling
perspective~\cite{ber21}.

In this work, we introduce hypothetical T$^\prime$-La$_2$NiO$_4$ and provide a
basic characterization of its theoretical correlated electronic structure. It is found that
the impact of the O$(2p)$ orbitals is much more pronounced compared to their role in
the T structure, resulting
in a largely-reduced charge-transfer energy and correspondingly, a much smaller charge gap in
the perfectly-ordered T$^\prime$ compound. For electron doping, a cuprate-like interacting Fermi
surface is revealed, yet with possibly a delicate role of additional Ni-$d_{z^2}$ derived features.

\textit{Theoretical approach.---}
First-principles calculations are performed on the level of DFT, as
well as via its charge-selfconsistent combination with dynamical mean-field theory (DMFT)
by including additional correlation effects on oxygen through self-interaction correction
(SIC) in the so-called DFT+sicDMFT approach~\cite{lec19}. The DFT part is here given by a
mixed-basis pseudopotential framework~\cite{elsaesser90,lechermann02,mbpp_code} in the
local-density approximation with a plane-wave cutoff $E_{\rm cut}=16$\,Ryd, and atomic-like
localized functions for La$(5d)$,
Ni$(3d)$ as well as O$(2s,2p)$. A $k$-point mesh tiling of $11$$\times$$11$$\times$$11$ is introduced.
Application of the SIC takes place via the O$(2s,2p)$ orbitals through weight factors $w_p$.
While the $2s$ orbital is
fully corrected with $w_p=1.0$, we choose $w_p=0.8$ for the $2p$ orbitals~\cite{korner10,lec19}.
Continuous-time quantum Monte Carlo in the hybridization-expansion scheme~\cite{werner06}
as implemented in the TRIQS code~\cite{parcollet15,seth16} solves the DMFT
quantum-impurity problem.
A five-orbital Slater-Hamiltonian, parameterized by Hubbard $U$ and Hund exchange $J$
governs the correlated subspace as defined by Ni$(3d)$ projected-local orbitals~\cite{amadon08}.
We use 19 Kohn-Sham bands  above the O$(2s)$ bands for this Wannier-like projection.
Throughout the calculations, the choice $J=1$\,eV for nickelates~\cite{lec22} is utilized.
A double-counting correction of the fully-localized-limit form~\cite{anisimov93} is applied.
The system temperature is set to $T=200\,$K in the calculations.

\begin{figure}[t]
      \includegraphics[width=\linewidth]{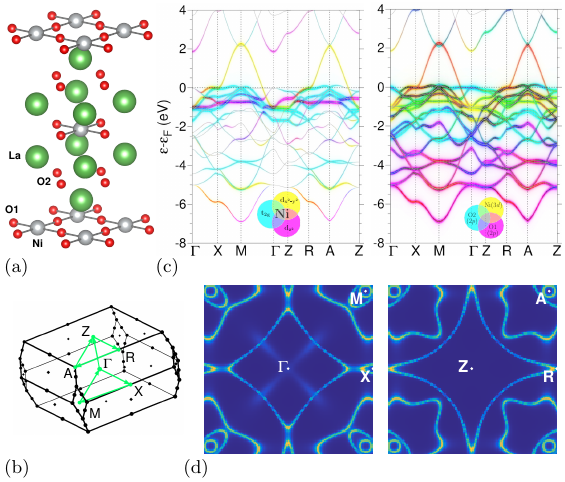}
      \caption{Basic characterization of tetragonal T$^\prime$-La$_2$NiO$_4$.
        (a) Crystal structure with La (green), Ni (grey)
        as well as O1 and O2 (both red) sites. (b) Brillouin zone (BZ) with path along
        high-symmetry lines. (c) DFT band structure in two fatspec representations, with
        spectral weights according to (left panel) Ni-$d_{x^2-y^2}$ (yellow), Ni-$d_{z^2}$
        (pink) and Ni-$t_{2g}$ (cyan); (right panel) according to Ni$(3d)$
        (yellow), O1$(2p)$ (pink) and O2$(2p)$ (cyan). (d) DFT Fermi surface in 
        $k_z=0$ (left panel) and $k_z=1/2$ (right panel) plane.\label{fig1}}
\end{figure}

\textit{Stoichiometric system.---}
The T$^\prime$-La$_2$NiO$_4$ compound is described by assuming tetragonal symmetry (see
Fig.~\ref{fig1}a). Compared to the T structure, the original apical-oxygen atoms are shifted
towards the O2 positions within the rocksalt block. The O1 atoms remain in the plane with Ni.
We optimized the lattice constants as well as the atomic positions within DFT using the
generalized-gradient approximation. As a result, the in-plane lattice constant increases
to $a=3.97$\,\AA\, compared to the T-structure value $a=3.88$\,\AA~\cite{rod91}, whereas the
ratio $c/a=1.61$ is not largely affected. A significant increase of the in-plane lattice
constant has also been reported for T$^\prime$-La$_2$CuO$_4$, which has been used
for successful thin-film growth on a tailored substrate~\cite{wei16}.

The DFT band structure is shown in Fig.~\ref{fig1}c by employing two kinds of fatspec representations,
i.e. Ni$(3d)$ orbitals (left panel), as well as Ni$(3d)$ shell against O1,2$(2p)$ (right panel),
in order to highlight the various site/orbital contributions. Overall, the result displays a
dominant Ni-$d_{x^2-y^2}$-O$(2p)$ dispersion around the Fermi level $\varepsilon_{\rm F}$,
yet with sizeable Ni-$d_{z^2}$ hybridization close to the X and R points in the BZ.
This wide dispersion forms a dominant hole Fermi-surface sheet
around the $\Gamma$ and Z point, with seemingly quite significant $k_z$ dispersion
(see Fig.~\ref{fig1}d). A prominent Ni-$d_{z^2}$ dispersion is visible around $-1$\,eV, and
the Ni-$t_{2g}$ weight also remains more or less exclusively below the Fermi level. The O$(2p)$
weights are quite interesting. Expectedly for O1 major contributions arise deep down in
energy for the bonding states with Ni$(3d)$ and above the Fermi level for the antibonding states
especially with Ni-$d_{x^2-y^2}$. However for O2, there is major contribution not too far below
and, surprisingly, very close to $\varepsilon_{\rm F}$. For instance, the flat-band part right
at the Fermi level along the $\Gamma$-M direction is dominantly stemming from the $2p$ states of
O2. Majorly O2-derived bands thus also contribute to the DFT Fermi surface with hole-pocket
structures around the M and A point (see Fig.~\ref{fig1}d). This substantial O2 low(er)-energy
spectral weight leads to a quite small charge-transfer energy $\Delta=1.9$\,eV, about half the value
associated with T-La$_2$NiO$_4$. The local DFT occupations for the formal $(d^8,p^6)$ compound
amount to
$\{n_{z^2},n_{x^2-y^2},n_{t_{2g}}\}=\{1.87,1.01,5.75\}$ for Ni$(3d)$ with $n_{3d}=8.63$, and an average
$n_{2p}=5.65$ for the oxygen sites.

\begin{figure}[b]
      \includegraphics[width=\linewidth]{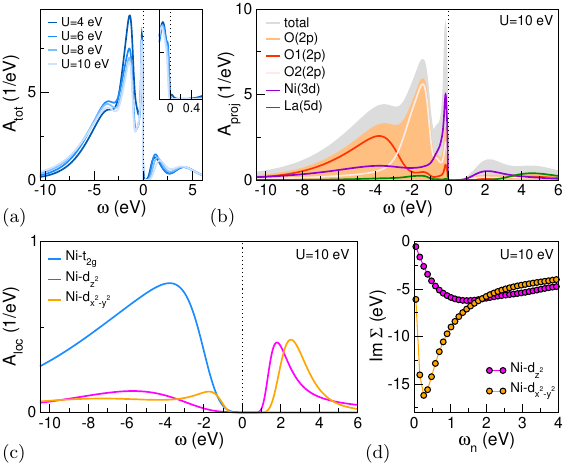}
      \caption{Spectral data for T$^\prime$-La$_2$NiO$_4$ from DFT+sicDMFT.
        (a) $U$-dependent total spectral function (inset: low-energy blow up).
        (b-d) Results for $U=10\,$eV: (b) total and projected $k$-integrated spectral function,
        (c) local Ni$(3d)$ spectral function, and (d) imaginary part of
        the Ni-$e_g$ self-energy in Matsubara space.\label{fig2}}
\end{figure}

When including electron-electron interactions on the DFT+sicDMFT level beyond the effective
single-particle paradigm, the T'-La$_2$NiO$_4$ compound becomes a correlated insulator, as documented
in Fig.~\ref{fig2}. From the $k$-integrated total spectral function $A_{\rm tot}(\omega)$
shown in Fig.~\ref{fig2}a, this happens already for moderate values $U\gtrsim 4$\,eV.
Importantly, the charge gap $\sim 1.8$\,eV is comparable to $\Delta$ and remains nearly
unchanged with increasing $U$, an obvious signature of a manifest charge-transfer
insulator~\cite{zaa85}. Note that when neglecting the SIC on oxygen, opening
a charge gap within standard DFT+DMFT appears rather challenging, pointing to the particularly
relevant role of the (correlated) ligand states in this compound.
\begin{figure*}[t]
      \includegraphics[width=\linewidth]{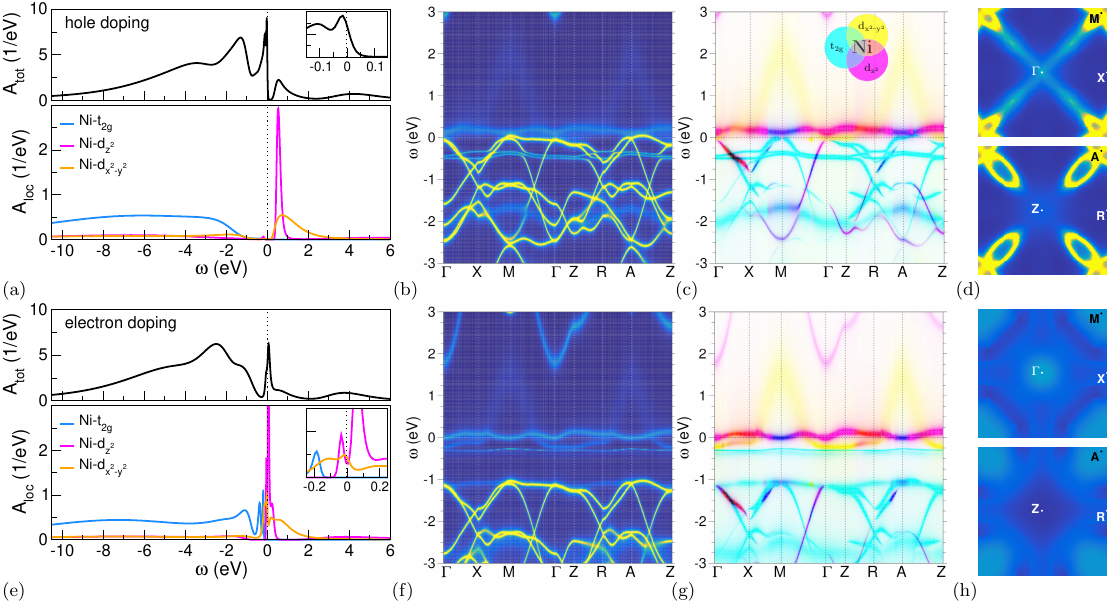}
      \caption{Spectral data for doped T$^\prime$-La$_2$NiO$_4$ 
        (doping level $\delta=0.2$) from DFT+sicDMFT with $U=10\,$eV: (a-d) hole doping,
        (e-h) electron doping.
        (a,e) Total spectral function with low-energy blow up as inset (upper panel) and local
        Ni$(3d)$ spectral function (lower panel). (b,f) $k$-resolved spectral function along
        high-symmetry lines, and (c,g) same using Ni$(3d)$ fatspec representation.
        (d,h) Interacting Fermi surface in $k_z$=0 plane (upper panel) and $k_z=1/2$ plane
        (lower panel).
        \label{fig3}}
\end{figure*}	
In order to keep a coherent setting for the local Coulomb interaction across the
nickelate series within DFT+sicDMFT~\cite{lec19,lec22,lec23}, we fix in the following a Hubbard
$U=10$\,eV. Figure~\ref{fig2}b displays a two-peak structure for the O$(2p)$ states in the correlated
spectrum, with a broader peak at $\sim -3.9$\,eV for O1 and a sharper peak at $\sim -1.7$\,eV
from O2. Significant O$(2p)$ weight remains close to the Fermi level also with correlations,
underlining the charge-transfer nature. Explicit La$(5d)$-derived states are seemingly of overall
minor relevance. Interestingly, while the projected Ni$(3d)$ spectral function peaks close to the
lower gap edge, there is no such peak in the local Ni$(3d)$ spectrum (cf. Fig.~\ref{fig2}c).
This may be understood again from the crucial impact of the ligand states, which shift $3d$
spectral weight via hybridization to the gap edge in the projected spectrum. The local
Ni$(3d)$ filling reads now $\{n_{z^2},n_{x^2-y^2},n_{t_{2g}}\}=\{1.01,1.11,5.99\}$, hence there is
especially relevant charge depletion of Ni-$d_{z^2}$ in the correlated regime. 
A similar correlation-induced Ni-$d_{z^2}$ charge depletion is observed in the akin bilayer compound
La$_3$Ni$_2$O$_6$. But whereas there the bilayer setting supports a very intriguing correlated
orbital-selective scenario~\cite{lec25}, it leads for the present single-layer compound to a robust
effective two-orbital charge-transfer insulator. Still, the leading Mott-critical orbital is
the Ni-$d_{x^2-y^2}$ one and the Ni-$d_{z^2}$ orbital is drawn into the insulating state, as suggested
from the orbital-asymmetric imaginary parts of the Matsubara self-energy $\Sigma(i\omega_n)$ in
Fig.~\ref{fig2}d. The O$(2p)$ filling is marginally reduced with correlations to $n_{2p}=5.58$,
i.e. stabilizing the sizeable ligand-hole content, but with no striking O1,O2 differentiation.

\textit{Doped regime.---}
Charge doping proves effective in metallizing the system, yet with a strong asymmetry concerning the
doped carrier type, as displayed in Fig.~\ref{fig3} for a charge-doping level $\delta=0.2$. In general,
doping appears to enhance the Ni-$e_g$ upper-Hubbard-band differentiation, leading to rather narrow
incoherent Ni-$d_{z^2}$ bands (see Figs.~\ref{fig3}a,e and c,g).
Upon hole doping (cf. Figs.~\ref{fig3}a-d),
the local Ni$(3d)$ shell remains insulating and itinerant carriers enter the low-lying mostly O2-derived
states. While the Ni-$e_g$ upper Hubbard bands still remain close to the Fermi level, they do not
participate in transport. As visible from Fig.~\ref{fig3}d, the hole-doped interacting Fermi surface
exhibits unusual character with electron-pocket structures close to M,A points and very narrow sheets
along $\Gamma$-M. Electron doping activates the local Ni$(3d)$ shell, and while the Ni-$t_{2g}$ states
remain fully occupied, the renormalized effective Ni-$d_{x^2-y^2}$ dispersion dominates the itinerant regime
(see Figs.~\ref{fig3}e-h). The Ni-$d_{z^2}$ states may play an intriguing role, since the corresponding
upper Hubbard band becomes also weakly doped, but with the Fermi level located in a pseudogap
(see inset of Fig.~\ref{fig3}e). Note that the local Ni$(3d)$ filling reads
$\{n_{z^2},n_{x^2-y^2},n_{t_{2g}}\}=\{1.06,1.25,5.98\}$ for electron doping, thus major doping
indeed takes place into Ni-$d_{x^2-y^2}$. Correspondingly, the electron-doped interacting Fermi surface
shows the well-known cuprate-like hole sheet around $\Gamma$ with weak dispersion along $k_z$ (cf.
Fig.~\ref{fig3}h). Minor spectral weight close to $\Gamma$ and around M,A points stems from
the Ni-$d_{z^2}$ contributions. While the Ni-$d_{x^2-y^2}$ features are computationally robust,
resolving the Ni-$d_{z^2}$ behavior upon electron doping is rather delicate, i.e. minor variations in the
Coulomb-interaction strengths and/or the setting of the projection window for the correlated subspace
may affect their (non)doping character.

\textit{Stability of NN-AFM ordering.---}
\begin{figure}[t]
      \includegraphics[width=\linewidth]{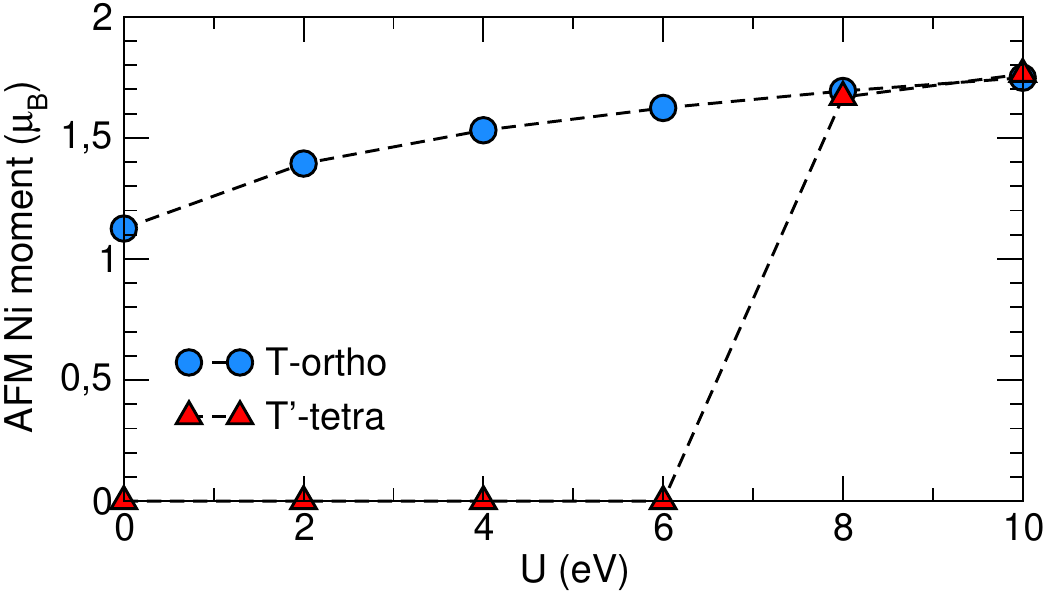}
      \caption{Nearest-neighbor antiferromagnetic-ordered Ni moment within DFT+U for La$_2$NiO$_4$
        in the orthorhombic T (blue color) and the tetragonal T$^\prime$ (red color) structure.
        \label{fig4}}
\end{figure}	
Let us finally comment on the possibility of antiferromagnetic ordering of stoichiometric 
T$^\prime$-La$_2$NiO$_4$. In the T structure, the system displays robust nearest-neighbor in-plane
AFM ordering~\cite{lan89,rod91}, which is well reproduced within the DFT+sicDMFT
formalism~\cite{lec22}, and
we restrict in the following the discussion to this most-simple spin-ordering pattern of AFM kind.
First, the effect of static correlations is examined by performing DFT+U
calculations, and the results for increasing Hubbard $U$ are summarized for the ordered AFM
Ni moment in Fig.~\ref{fig4}. The Ni moment turns out robust even in the DFT limit of vanishing
$U$ for the orthorhombic T structure. For comparison, we also computed the AFM moment in the
(high-temperature) tetragonal T structure, but the values were only marginally smaller.
On the other hand, for the T$^\prime$ structure the AFM ordering
tendency is strongly reduced, resulting only for $U>6$\,eV in a stable Ni moment.
A ferromagnetic solution is stabilized for smaller $U$. 

In order to check these (meta)stability findings against the impact of local quantum fluctuations,
we also conducted AFM DFT+sicDMFT calculations in the intermediate-coupling scheme~\cite{lec22}.
The latter handles the ordered spin polarization within the DFT+sic part only on the (nonlocal)
plane-wave level, whereas the (local) atomic-like function contribution is spin averaged. This ensures
that the spin polarization is originating from the local DMFT self-energy, while still keeping
a (nonlocal) spin coupling to the remaining Hilbert space. Note that a complete spin polarization
also within the DFT part often artificially enhances magnetic-ordering tendencies. On the other hand,
a sole DMFT
spin-dependent self-energy without any spin polarization in the DFT part often misses contributions
from ligand-mediated exchange mechanisms and may result in too weak magnetic-ordering tendencies.
The IC scheme produces strong AFM Ni moments for T-La$_2$NiO$_4$ in very good accordance with
experiment~\cite{lec22}. For T$^\prime$-La$_2$NiO$_4$ however, those moments vanish
($m_{\rm Ni}<0.1\,\mu_{\rm B}$) for $U=10$\,eV at $T=200$\,K, and even so for a lower $T=100$\,K.
Thus, while static DFT+U calculation still see AFM ordering at larger $U$, additional quantum
fluctuations within DFT+sicDMFT either lead to a rather small N{\'e}el temperature, or destroy such an
order altogether. Apparently, the missing apical oxygen near the Ni sites are effective in
suppressing AFM order, as already observed in several reduced nickelate
compounds~\cite{hay99,apR13,zhang17,pan21,lec21}. But then note that in some cases~\cite{ande07,beh15},
defects might again foster magnetic ordering tendencies, hence a more thorough theoretical study of
this issue should take place once experimental work succeeds in stabilizing the T$^\prime$ structure.

\textit{Discussion.---}
The T$^\prime$-La$_2$NiO$_4$ compound of the RP nickelate series has
so far not been stabilized in experiment. However from a theoretical point of view, its
correlated electronic structure displays some highly interesting features. Different from
the known single-layer T structure nickelate, the particularly strong charge-transfer
physics gives way to a decisive role of the O$(2p)$ orbitals in the formation of a
correlation-induced insulating state at stoichiometry. While hole doping enables
intriguing low-energy O$(2p)$-derived states, electron doping puts this T$^\prime$
compound in rather strong connection to doped cuprates, despite a possible multiorbital
$3d$ scenario from additional Ni-$d_{z^2}$ correlated states. In reality, additional
defect formation might lead to electron-doping characteristics already close to stoichiometry,
similarly as in single-layer T$^\prime$ cuprates~\cite{wei16}. For an experimental stabilization
of T$^\prime$-La$_2$NiO$_4$, thin-film growth of single-layer RP nickelate under tensile
strain appears most promising. Last but not least, though
challenging experimentally, intensifying the nickelate research towards the minimal-layer
($n=1,2$) T$^\prime$-RP compounds may provide important insight in the complex interplay
between Ni$(3d)$ and O$(2p)$ orbitals. A deeper understanding of nickelate superconductivity
and its relation to the cuprate case may depend on such insights.

\section{Acknowledgements} 
The work is supported by the German Research Foundation within the Project Number 572794210. 
Computations were performed at the Ruhr-University Bochum.

\bibliography{literatur}

\begin{thebibliography}{46}%
\makeatletter
\providecommand \@ifxundefined [1]{%
 \@ifx{#1\undefined}
}%
\providecommand \@ifnum [1]{%
 \ifnum #1\expandafter \@firstoftwo
 \else \expandafter \@secondoftwo
 \fi
}%
\providecommand \@ifx [1]{%
 \ifx #1\expandafter \@firstoftwo
 \else \expandafter \@secondoftwo
 \fi
}%
\providecommand \natexlab [1]{#1}%
\providecommand \enquote  [1]{``#1''}%
\providecommand \bibnamefont  [1]{#1}%
\providecommand \bibfnamefont [1]{#1}%
\providecommand \citenamefont [1]{#1}%
\providecommand \href@noop [0]{\@secondoftwo}%
\providecommand \href [0]{\begingroup \@sanitize@url \@href}%
\providecommand \@href[1]{\@@startlink{#1}\@@href}%
\providecommand \@@href[1]{\endgroup#1\@@endlink}%
\providecommand \@sanitize@url [0]{\catcode `\\12\catcode `\$12\catcode
  `\&12\catcode `\#12\catcode `\^12\catcode `\_12\catcode `\%12\relax}%
\providecommand \@@startlink[1]{}%
\providecommand \@@endlink[0]{}%
\providecommand \url  [0]{\begingroup\@sanitize@url \@url }%
\providecommand \@url [1]{\endgroup\@href {#1}{\urlprefix }}%
\providecommand \urlprefix  [0]{URL }%
\providecommand \Eprint [0]{\href }%
\providecommand \doibase [0]{https://doi.org/}%
\providecommand \selectlanguage [0]{\@gobble}%
\providecommand \bibinfo  [0]{\@secondoftwo}%
\providecommand \bibfield  [0]{\@secondoftwo}%
\providecommand \translation [1]{[#1]}%
\providecommand \BibitemOpen [0]{}%
\providecommand \bibitemStop [0]{}%
\providecommand \bibitemNoStop [0]{.\EOS\space}%
\providecommand \EOS [0]{\spacefactor3000\relax}%
\providecommand \BibitemShut  [1]{\csname bibitem#1\endcsname}%
\let\auto@bib@innerbib\@empty
\bibitem [{\citenamefont {Goodenough}(1973)}]{goo73}%
  \BibitemOpen
  \bibfield  {author} {\bibinfo {author} {\bibfnamefont {J.}~\bibnamefont
  {Goodenough}},\ }\bibfield  {title} {\bibinfo {title} {Interpretation of the
  transport properties of ln2nio4 and ln2cuo4 compounds},\ }\href
  {https://doi.org/https://doi.org/10.1016/0025-5408(73)90046-9} {\bibfield
  {journal} {\bibinfo  {journal} {Materials Research Bulletin}\ }\textbf
  {\bibinfo {volume} {8}},\ \bibinfo {pages} {423} (\bibinfo {year}
  {1973})}\BibitemShut {NoStop}%
\bibitem [{\citenamefont {Imada}\ \emph {et~al.}(1998)\citenamefont {Imada},
  \citenamefont {Fujimori},\ and\ \citenamefont {Tokura}}]{ima98}%
  \BibitemOpen
  \bibfield  {author} {\bibinfo {author} {\bibfnamefont {M.}~\bibnamefont
  {Imada}}, \bibinfo {author} {\bibfnamefont {A.}~\bibnamefont {Fujimori}},\
  and\ \bibinfo {author} {\bibfnamefont {Y.}~\bibnamefont {Tokura}},\
  }\bibfield  {title} {\bibinfo {title} {Metal-insulator transitions},\ }\href
  {https://doi.org/10.1103/RevModPhys.70.1039} {\bibfield  {journal} {\bibinfo
  {journal} {Rev. Mod. Phys.}\ }\textbf {\bibinfo {volume} {70}},\ \bibinfo
  {pages} {1039} (\bibinfo {year} {1998})}\BibitemShut {NoStop}%
\bibitem [{\citenamefont {Tokura}\ \emph {et~al.}(1989)\citenamefont {Tokura},
  \citenamefont {Takagi},\ and\ \citenamefont {Uchida}}]{tok89}%
  \BibitemOpen
  \bibfield  {author} {\bibinfo {author} {\bibfnamefont {Y.}~\bibnamefont
  {Tokura}}, \bibinfo {author} {\bibfnamefont {H.}~\bibnamefont {Takagi}},\
  and\ \bibinfo {author} {\bibfnamefont {S.}~\bibnamefont {Uchida}},\
  }\bibfield  {title} {\bibinfo {title} {A superconducting copper oxide
  compound with electrons as the charge carriers},\ }\href
  {https://doi.org/10.1038/337345a0} {\bibfield  {journal} {\bibinfo  {journal}
  {Nature}\ }\textbf {\bibinfo {volume} {337}},\ \bibinfo {pages} {345}
  (\bibinfo {year} {1989})}\BibitemShut {NoStop}%
\bibitem [{\citenamefont {Chou}\ \emph {et~al.}(1990)\citenamefont {Chou},
  \citenamefont {Cho}, \citenamefont {Miller},\ and\ \citenamefont
  {Johnston}}]{cho90}%
  \BibitemOpen
  \bibfield  {author} {\bibinfo {author} {\bibfnamefont {F.~C.}\ \bibnamefont
  {Chou}}, \bibinfo {author} {\bibfnamefont {J.~H.}\ \bibnamefont {Cho}},
  \bibinfo {author} {\bibfnamefont {L.~L.}\ \bibnamefont {Miller}},\ and\
  \bibinfo {author} {\bibfnamefont {D.~C.}\ \bibnamefont {Johnston}},\
  }\bibfield  {title} {\bibinfo {title} {New phases induced by hydrogen
  reduction and by subsequent oxidation of
  ${\mathit{l}}_{2}$${\mathrm{cuo}}_{4}$ (l=la,pr,nd,sm,eu,gd)},\ }\href
  {https://doi.org/10.1103/PhysRevB.42.6172} {\bibfield  {journal} {\bibinfo
  {journal} {Phys. Rev. B}\ }\textbf {\bibinfo {volume} {42}},\ \bibinfo
  {pages} {6172} (\bibinfo {year} {1990})}\BibitemShut {NoStop}%
\bibitem [{\citenamefont {Matsumoto}\ \emph {et~al.}(2009)\citenamefont
  {Matsumoto}, \citenamefont {Utsuki}, \citenamefont {Tsukada}, \citenamefont
  {Yamamoto}, \citenamefont {Manabe},\ and\ \citenamefont {Naito}}]{mat09}%
  \BibitemOpen
  \bibfield  {author} {\bibinfo {author} {\bibfnamefont {O.}~\bibnamefont
  {Matsumoto}}, \bibinfo {author} {\bibfnamefont {A.}~\bibnamefont {Utsuki}},
  \bibinfo {author} {\bibfnamefont {A.}~\bibnamefont {Tsukada}}, \bibinfo
  {author} {\bibfnamefont {H.}~\bibnamefont {Yamamoto}}, \bibinfo {author}
  {\bibfnamefont {T.}~\bibnamefont {Manabe}},\ and\ \bibinfo {author}
  {\bibfnamefont {M.}~\bibnamefont {Naito}},\ }\bibfield  {title} {\bibinfo
  {title} {Synthesis and properties of superconducting
  ${T}^{\ensuremath{'}}\text{\ensuremath{-}}{R}_{2}{\text{cuo}}_{4}$
  ($r=\text{Pr}$, nd, sm, eu, gd)},\ }\href
  {https://doi.org/10.1103/PhysRevB.79.100508} {\bibfield  {journal} {\bibinfo
  {journal} {Phys. Rev. B}\ }\textbf {\bibinfo {volume} {79}},\ \bibinfo
  {pages} {100508(R)} (\bibinfo {year} {2009})}\BibitemShut {NoStop}%
\bibitem [{\citenamefont {Brinkmann}\ \emph {et~al.}(1995)\citenamefont
  {Brinkmann}, \citenamefont {Rex}, \citenamefont {Bach},\ and\ \citenamefont
  {Westerholt}}]{bri95}%
  \BibitemOpen
  \bibfield  {author} {\bibinfo {author} {\bibfnamefont {M.}~\bibnamefont
  {Brinkmann}}, \bibinfo {author} {\bibfnamefont {T.}~\bibnamefont {Rex}},
  \bibinfo {author} {\bibfnamefont {H.}~\bibnamefont {Bach}},\ and\ \bibinfo
  {author} {\bibfnamefont {K.}~\bibnamefont {Westerholt}},\ }\bibfield  {title}
  {\bibinfo {title} {Extended superconducting concentration range observed in
  ${\mathrm{pr}}_{2\ensuremath{-}\mathit{x}}{\mathrm{ce}}_{\mathit{x}}{\mathrm{cuo}}_{4\ensuremath{-}\ensuremath{\delta}}$},\
  }\href {https://doi.org/10.1103/PhysRevLett.74.4927} {\bibfield  {journal}
  {\bibinfo  {journal} {Phys. Rev. Lett.}\ }\textbf {\bibinfo {volume} {74}},\
  \bibinfo {pages} {4927} (\bibinfo {year} {1995})}\BibitemShut {NoStop}%
\bibitem [{\citenamefont {Sekitani}\ \emph {et~al.}(2003)\citenamefont
  {Sekitani}, \citenamefont {Naito},\ and\ \citenamefont {Miura}}]{sek03}%
  \BibitemOpen
  \bibfield  {author} {\bibinfo {author} {\bibfnamefont {T.}~\bibnamefont
  {Sekitani}}, \bibinfo {author} {\bibfnamefont {M.}~\bibnamefont {Naito}},\
  and\ \bibinfo {author} {\bibfnamefont {N.}~\bibnamefont {Miura}},\ }\bibfield
   {title} {\bibinfo {title} {Kondo effect in underdoped n-type
  superconductors},\ }\href {https://doi.org/10.1103/PhysRevB.67.174503}
  {\bibfield  {journal} {\bibinfo  {journal} {Phys. Rev. B}\ }\textbf {\bibinfo
  {volume} {67}},\ \bibinfo {pages} {174503} (\bibinfo {year}
  {2003})}\BibitemShut {NoStop}%
\bibitem [{\citenamefont {Wei}\ \emph {et~al.}(2016)\citenamefont {Wei},
  \citenamefont {Adamo}, \citenamefont {Nowadnick}, \citenamefont {Lochocki},
  \citenamefont {Chatterjee}, \citenamefont {Ruf}, \citenamefont {Beasley},
  \citenamefont {Schlom},\ and\ \citenamefont {Shen}}]{wei16}%
  \BibitemOpen
  \bibfield  {author} {\bibinfo {author} {\bibfnamefont {H.~I.}\ \bibnamefont
  {Wei}}, \bibinfo {author} {\bibfnamefont {C.}~\bibnamefont {Adamo}}, \bibinfo
  {author} {\bibfnamefont {E.~A.}\ \bibnamefont {Nowadnick}}, \bibinfo {author}
  {\bibfnamefont {E.~B.}\ \bibnamefont {Lochocki}}, \bibinfo {author}
  {\bibfnamefont {S.}~\bibnamefont {Chatterjee}}, \bibinfo {author}
  {\bibfnamefont {J.~P.}\ \bibnamefont {Ruf}}, \bibinfo {author} {\bibfnamefont
  {M.~R.}\ \bibnamefont {Beasley}}, \bibinfo {author} {\bibfnamefont {D.~G.}\
  \bibnamefont {Schlom}},\ and\ \bibinfo {author} {\bibfnamefont {K.~M.}\
  \bibnamefont {Shen}},\ }\bibfield  {title} {\bibinfo {title} {Electron doping
  of the parent cuprate ${\mathrm{la}}_{2}{\mathrm{cuo}}_{4}$ without cation
  substitution},\ }\href {https://doi.org/10.1103/PhysRevLett.117.147002}
  {\bibfield  {journal} {\bibinfo  {journal} {Phys. Rev. Lett.}\ }\textbf
  {\bibinfo {volume} {117}},\ \bibinfo {pages} {147002} (\bibinfo {year}
  {2016})}\BibitemShut {NoStop}%
\bibitem [{\citenamefont {Li}\ \emph {et~al.}(2019)\citenamefont {Li},
  \citenamefont {Lee}, \citenamefont {Wang}, \citenamefont {Osada},
  \citenamefont {Crossley}, \citenamefont {Lee}, \citenamefont {Cui},
  \citenamefont {Hikita},\ and\ \citenamefont {Hwang}}]{li19}%
  \BibitemOpen
  \bibfield  {author} {\bibinfo {author} {\bibfnamefont {D.}~\bibnamefont
  {Li}}, \bibinfo {author} {\bibfnamefont {K.}~\bibnamefont {Lee}}, \bibinfo
  {author} {\bibfnamefont {B.~Y.}\ \bibnamefont {Wang}}, \bibinfo {author}
  {\bibfnamefont {M.}~\bibnamefont {Osada}}, \bibinfo {author} {\bibfnamefont
  {S.}~\bibnamefont {Crossley}}, \bibinfo {author} {\bibfnamefont {H.~R.}\
  \bibnamefont {Lee}}, \bibinfo {author} {\bibfnamefont {Y.}~\bibnamefont
  {Cui}}, \bibinfo {author} {\bibfnamefont {Y.}~\bibnamefont {Hikita}},\ and\
  \bibinfo {author} {\bibfnamefont {H.~Y.}\ \bibnamefont {Hwang}},\ }\bibfield
  {title} {\bibinfo {title} {Superconductivity in an infinite-layer
  nickelate},\ }\href {https://doi.org/10.1038/s41586-019-1496-5} {\bibfield
  {journal} {\bibinfo  {journal} {Nature}\ }\textbf {\bibinfo {volume} {572}},\
  \bibinfo {pages} {624} (\bibinfo {year} {2019})}\BibitemShut {NoStop}%
\bibitem [{\citenamefont {Sun}\ \emph {et~al.}(2023)\citenamefont {Sun},
  \citenamefont {Huo}, \citenamefont {Hu}, \citenamefont {Li}, \citenamefont
  {Han}, \citenamefont {Tang}, \citenamefont {Mao}, \citenamefont {Yang},
  \citenamefont {Wang}, \citenamefont {Cheng}, \citenamefont {Yao},
  \citenamefont {Zhang},\ and\ \citenamefont {Wang}}]{sun23}%
  \BibitemOpen
  \bibfield  {author} {\bibinfo {author} {\bibfnamefont {H.}~\bibnamefont
  {Sun}}, \bibinfo {author} {\bibfnamefont {M.}~\bibnamefont {Huo}}, \bibinfo
  {author} {\bibfnamefont {X.}~\bibnamefont {Hu}}, \bibinfo {author}
  {\bibfnamefont {J.}~\bibnamefont {Li}}, \bibinfo {author} {\bibfnamefont
  {Y.}~\bibnamefont {Han}}, \bibinfo {author} {\bibfnamefont {L.}~\bibnamefont
  {Tang}}, \bibinfo {author} {\bibfnamefont {Z.}~\bibnamefont {Mao}}, \bibinfo
  {author} {\bibfnamefont {P.}~\bibnamefont {Yang}}, \bibinfo {author}
  {\bibfnamefont {B.}~\bibnamefont {Wang}}, \bibinfo {author} {\bibfnamefont
  {J.}~\bibnamefont {Cheng}}, \bibinfo {author} {\bibfnamefont {D.-X.}\
  \bibnamefont {Yao}}, \bibinfo {author} {\bibfnamefont {G.-M.}\ \bibnamefont
  {Zhang}},\ and\ \bibinfo {author} {\bibfnamefont {M.}~\bibnamefont {Wang}},\
  }\bibfield  {title} {\bibinfo {title} {Signatures of superconductivity near
  80 k in a nickelate under high pressure},\ }\href
  {https://doi.org/10.1038/s41586-023-06408-7} {\bibfield  {journal} {\bibinfo
  {journal} {Nature}\ }\textbf {\bibinfo {volume} {621}},\ \bibinfo {pages}
  {493} (\bibinfo {year} {2023})}\BibitemShut {NoStop}%
\bibitem [{\citenamefont {Lander}\ \emph {et~al.}(1989)\citenamefont {Lander},
  \citenamefont {Brown}, \citenamefont {Spal/ek},\ and\ \citenamefont
  {Honig}}]{lan89}%
  \BibitemOpen
  \bibfield  {author} {\bibinfo {author} {\bibfnamefont {G.~H.}\ \bibnamefont
  {Lander}}, \bibinfo {author} {\bibfnamefont {P.~J.}\ \bibnamefont {Brown}},
  \bibinfo {author} {\bibfnamefont {J.}~\bibnamefont {Spal/ek}},\ and\ \bibinfo
  {author} {\bibfnamefont {J.~M.}\ \bibnamefont {Honig}},\ }\bibfield  {title}
  {\bibinfo {title} {Structural and magnetization density studies of
  ${\mathrm{la}}_{2}$${\mathrm{nio}}_{4}$},\ }\href
  {https://doi.org/10.1103/PhysRevB.40.4463} {\bibfield  {journal} {\bibinfo
  {journal} {Phys. Rev. B}\ }\textbf {\bibinfo {volume} {40}},\ \bibinfo
  {pages} {4463} (\bibinfo {year} {1989})}\BibitemShut {NoStop}%
\bibitem [{\citenamefont {Rodriguez-Carvajal}\ \emph
  {et~al.}(1991)\citenamefont {Rodriguez-Carvajal}, \citenamefont
  {Fernandez-Diaz},\ and\ \citenamefont {Martinez}}]{rod91}%
  \BibitemOpen
  \bibfield  {author} {\bibinfo {author} {\bibfnamefont {J.}~\bibnamefont
  {Rodriguez-Carvajal}}, \bibinfo {author} {\bibfnamefont {M.~T.}\ \bibnamefont
  {Fernandez-Diaz}},\ and\ \bibinfo {author} {\bibfnamefont {J.~L.}\
  \bibnamefont {Martinez}},\ }\bibfield  {title} {\bibinfo {title} {Neutron
  diffraction study on structural and magnetic properties of la$_2$nio$_4$},\
  }\href {https://doi.org/10.1088/0953-8984/3/19/002} {\bibfield  {journal}
  {\bibinfo  {journal} {Journal of Physics: Condensed Matter}\ }\textbf
  {\bibinfo {volume} {3}},\ \bibinfo {pages} {3215} (\bibinfo {year}
  {1991})}\BibitemShut {NoStop}%
\bibitem [{\citenamefont {Eisaki}\ \emph {et~al.}(1992)\citenamefont {Eisaki},
  \citenamefont {Uchida}, \citenamefont {Mizokawa}, \citenamefont {Namatame},
  \citenamefont {Fujimori}, \citenamefont {van Elp}, \citenamefont {Kuiper},
  \citenamefont {Sawatzky}, \citenamefont {Hosoya},\ and\ \citenamefont
  {Katayama-Yoshida}}]{eis92}%
  \BibitemOpen
  \bibfield  {author} {\bibinfo {author} {\bibfnamefont {H.}~\bibnamefont
  {Eisaki}}, \bibinfo {author} {\bibfnamefont {S.}~\bibnamefont {Uchida}},
  \bibinfo {author} {\bibfnamefont {T.}~\bibnamefont {Mizokawa}}, \bibinfo
  {author} {\bibfnamefont {H.}~\bibnamefont {Namatame}}, \bibinfo {author}
  {\bibfnamefont {A.}~\bibnamefont {Fujimori}}, \bibinfo {author}
  {\bibfnamefont {J.}~\bibnamefont {van Elp}}, \bibinfo {author} {\bibfnamefont
  {P.}~\bibnamefont {Kuiper}}, \bibinfo {author} {\bibfnamefont {G.~A.}\
  \bibnamefont {Sawatzky}}, \bibinfo {author} {\bibfnamefont {S.}~\bibnamefont
  {Hosoya}},\ and\ \bibinfo {author} {\bibfnamefont {H.}~\bibnamefont
  {Katayama-Yoshida}},\ }\bibfield  {title} {\bibinfo {title} {Electronic
  structure of
  ${\mathrm{la}}_{2\mathrm{\ensuremath{-}}\mathit{x}}$${\mathrm{sr}}_{\mathit{x}}$${\mathrm{nio}}_{4}$
  studied by photoemission and inverse-photoemission spectroscopy},\ }\href
  {https://doi.org/10.1103/PhysRevB.45.12513} {\bibfield  {journal} {\bibinfo
  {journal} {Phys. Rev. B}\ }\textbf {\bibinfo {volume} {45}},\ \bibinfo
  {pages} {12513} (\bibinfo {year} {1992})}\BibitemShut {NoStop}%
\bibitem [{\citenamefont {Guo}\ and\ \citenamefont {Temmerman}(1988)}]{guo88}%
  \BibitemOpen
  \bibfield  {author} {\bibinfo {author} {\bibfnamefont {G.~Y.}\ \bibnamefont
  {Guo}}\ and\ \bibinfo {author} {\bibfnamefont {W.~M.}\ \bibnamefont
  {Temmerman}},\ }\bibfield  {title} {\bibinfo {title} {Electronic structure
  and magnetism in la$_2$nio$_4$},\ }\href
  {https://doi.org/10.1088/0022-3719/21/22/006} {\bibfield  {journal} {\bibinfo
   {journal} {J. Phys. C}\ }\textbf {\bibinfo {volume} {21}},\ \bibinfo {pages}
  {L917} (\bibinfo {year} {1988})}\BibitemShut {NoStop}%
\bibitem [{\citenamefont {Pardo}\ \emph {et~al.}(2012)\citenamefont {Pardo},
  \citenamefont {Botana},\ and\ \citenamefont {Baldomir}}]{par12}%
  \BibitemOpen
  \bibfield  {author} {\bibinfo {author} {\bibfnamefont {V.}~\bibnamefont
  {Pardo}}, \bibinfo {author} {\bibfnamefont {A.~S.}\ \bibnamefont {Botana}},\
  and\ \bibinfo {author} {\bibfnamefont {D.}~\bibnamefont {Baldomir}},\
  }\bibfield  {title} {\bibinfo {title} {Enhanced thermoelectric response of
  hole-doped la${}_{2}$nio${}_{4+\ensuremath{\delta}}$ from ab initio
  calculations},\ }\href {https://doi.org/10.1103/PhysRevB.86.165114}
  {\bibfield  {journal} {\bibinfo  {journal} {Phys. Rev. B}\ }\textbf {\bibinfo
  {volume} {86}},\ \bibinfo {pages} {165114} (\bibinfo {year}
  {2012})}\BibitemShut {NoStop}%
\bibitem [{\citenamefont {Lane}\ and\ \citenamefont {Zhu}(2020)}]{lan20}%
  \BibitemOpen
  \bibfield  {author} {\bibinfo {author} {\bibfnamefont {C.}~\bibnamefont
  {Lane}}\ and\ \bibinfo {author} {\bibfnamefont {J.-X.}\ \bibnamefont {Zhu}},\
  }\bibfield  {title} {\bibinfo {title} {Landscape of coexisting excitonic
  states in the insulating single-layer cuprates and nickelates},\ }\href
  {https://doi.org/10.1103/PhysRevB.101.155135} {\bibfield  {journal} {\bibinfo
   {journal} {Phys. Rev. B}\ }\textbf {\bibinfo {volume} {101}},\ \bibinfo
  {pages} {155135} (\bibinfo {year} {2020})}\BibitemShut {NoStop}%
\bibitem [{\citenamefont {Lechermann}(2022)}]{lec22}%
  \BibitemOpen
  \bibfield  {author} {\bibinfo {author} {\bibfnamefont {F.}~\bibnamefont
  {Lechermann}},\ }\bibfield  {title} {\bibinfo {title} {Assessing the
  correlated electronic structure of lanthanum nickelates},\ }\href
  {https://doi.org/10.1088/2516-1075/ac5c6a} {\bibfield  {journal} {\bibinfo
  {journal} {Electron. Struct.}\ }\textbf {\bibinfo {volume} {4}},\ \bibinfo
  {pages} {015005} (\bibinfo {year} {2022})}\BibitemShut {NoStop}%
\bibitem [{\citenamefont {LaBollita}\ \emph {et~al.}(2024)\citenamefont
  {LaBollita}, \citenamefont {Bag}, \citenamefont {Kapeghian},\ and\
  \citenamefont {Botana}}]{lab24}%
  \BibitemOpen
  \bibfield  {author} {\bibinfo {author} {\bibfnamefont {H.}~\bibnamefont
  {LaBollita}}, \bibinfo {author} {\bibfnamefont {S.}~\bibnamefont {Bag}},
  \bibinfo {author} {\bibfnamefont {J.}~\bibnamefont {Kapeghian}},\ and\
  \bibinfo {author} {\bibfnamefont {A.~S.}\ \bibnamefont {Botana}},\ }\bibfield
   {title} {\bibinfo {title} {Electronic correlations, layer distinction, and
  electron doping in the alternating single-layer--trilayer
  ${\mathrm{la}}_{3}{\mathrm{ni}}_{2}{\mathrm{o}}_{7}$ polymorph},\ }\href
  {https://doi.org/10.1103/PhysRevB.110.155145} {\bibfield  {journal} {\bibinfo
   {journal} {Phys. Rev. B}\ }\textbf {\bibinfo {volume} {110}},\ \bibinfo
  {pages} {155145} (\bibinfo {year} {2024})}\BibitemShut {NoStop}%
\bibitem [{\citenamefont {Tang}\ \emph {et~al.}(2025)\citenamefont {Tang},
  \citenamefont {Wang}, \citenamefont {Liu}, \citenamefont {Lu}, \citenamefont
  {Wang}, \citenamefont {Lin},\ and\ \citenamefont {Zou}}]{tang25}%
  \BibitemOpen
  \bibfield  {author} {\bibinfo {author} {\bibfnamefont {S.-H.}\ \bibnamefont
  {Tang}}, \bibinfo {author} {\bibfnamefont {H.-Y.}\ \bibnamefont {Wang}},
  \bibinfo {author} {\bibfnamefont {D.-Y.}\ \bibnamefont {Liu}}, \bibinfo
  {author} {\bibfnamefont {F.}~\bibnamefont {Lu}}, \bibinfo {author}
  {\bibfnamefont {W.-H.}\ \bibnamefont {Wang}}, \bibinfo {author}
  {\bibfnamefont {H.~Q.}\ \bibnamefont {Lin}},\ and\ \bibinfo {author}
  {\bibfnamefont {L.-J.}\ \bibnamefont {Zou}},\ }\href
  {https://arxiv.org/abs/2511.15486} {\bibinfo {title} {Evolution of correlated
  electronic states of la2nio4 under hydrostatic pressure}} (\bibinfo {year}
  {2025}),\ \Eprint {https://arxiv.org/abs/2511.15486} {arXiv:2511.15486
  [cond-mat.str-el]} \BibitemShut {NoStop}%
\bibitem [{\citenamefont {Takeda}\ \emph {et~al.}(1990)\citenamefont {Takeda},
  \citenamefont {Kanno}, \citenamefont {Sakano}, \citenamefont {Yamamoto},
  \citenamefont {Takano}, \citenamefont {Bando}, \citenamefont {Akinaga},
  \citenamefont {Takita},\ and\ \citenamefont {Goodenough}}]{tak90}%
  \BibitemOpen
  \bibfield  {author} {\bibinfo {author} {\bibfnamefont {Y.}~\bibnamefont
  {Takeda}}, \bibinfo {author} {\bibfnamefont {R.}~\bibnamefont {Kanno}},
  \bibinfo {author} {\bibfnamefont {M.}~\bibnamefont {Sakano}}, \bibinfo
  {author} {\bibfnamefont {O.}~\bibnamefont {Yamamoto}}, \bibinfo {author}
  {\bibfnamefont {M.}~\bibnamefont {Takano}}, \bibinfo {author} {\bibfnamefont
  {Y.}~\bibnamefont {Bando}}, \bibinfo {author} {\bibfnamefont
  {H.}~\bibnamefont {Akinaga}}, \bibinfo {author} {\bibfnamefont
  {K.}~\bibnamefont {Takita}},\ and\ \bibinfo {author} {\bibfnamefont
  {J.}~\bibnamefont {Goodenough}},\ }\bibfield  {title} {\bibinfo {title}
  {Crystal chemistry and physical properties of la$_{2-x}$sr$_x$nio$_4$},\
  }\href {https://doi.org/10.1016/0025-5408(90)90100-G} {\bibfield  {journal}
  {\bibinfo  {journal} {Mater. Res. Bull.}\ }\textbf {\bibinfo {volume} {25}},\
  \bibinfo {pages} {293} (\bibinfo {year} {1990})}\BibitemShut {NoStop}%
\bibitem [{\citenamefont {Sreedhar}\ and\ \citenamefont {Rao}(1990)}]{sre90}%
  \BibitemOpen
  \bibfield  {author} {\bibinfo {author} {\bibfnamefont {K.}~\bibnamefont
  {Sreedhar}}\ and\ \bibinfo {author} {\bibfnamefont {C.~N.~R.}\ \bibnamefont
  {Rao}},\ }\bibfield  {title} {\bibinfo {title} {Electrical and magnetic
  properties of la$_{2-x}$sr$_x$nio$_4$: A tentative phase diagram},\ }\href
  {https://doi.org/10.1016/0025-5408(90)90079-H} {\bibfield  {journal}
  {\bibinfo  {journal} {Mater. Res. Bull.}\ }\textbf {\bibinfo {volume} {25}},\
  \bibinfo {pages} {1235} (\bibinfo {year} {1990})}\BibitemShut {NoStop}%
\bibitem [{\citenamefont {Cava}\ \emph {et~al.}(1991)\citenamefont {Cava},
  \citenamefont {Batlogg}, \citenamefont {Palstra}, \citenamefont {Krajewski},
  \citenamefont {Peck}, \citenamefont {Ramirez},\ and\ \citenamefont
  {Rupp}}]{cav91}%
  \BibitemOpen
  \bibfield  {author} {\bibinfo {author} {\bibfnamefont {R.~J.}\ \bibnamefont
  {Cava}}, \bibinfo {author} {\bibfnamefont {B.}~\bibnamefont {Batlogg}},
  \bibinfo {author} {\bibfnamefont {T.~T.}\ \bibnamefont {Palstra}}, \bibinfo
  {author} {\bibfnamefont {J.~J.}\ \bibnamefont {Krajewski}}, \bibinfo {author}
  {\bibfnamefont {W.~F.}\ \bibnamefont {Peck}}, \bibinfo {author}
  {\bibfnamefont {A.~P.}\ \bibnamefont {Ramirez}},\ and\ \bibinfo {author}
  {\bibfnamefont {L.~W.}\ \bibnamefont {Rupp}},\ }\bibfield  {title} {\bibinfo
  {title} {Magnetic and electrical properties of
  ${\mathrm{la}}_{2\mathrm{\ensuremath{-}}\mathit{x}}$${\mathrm{sr}}_{\mathit{x}}$${\mathrm{nio}}_{4\ifmmode\pm\else\textpm\fi{}\mathrm{\ensuremath{\delta}}}$},\
  }\href {https://doi.org/10.1103/PhysRevB.43.1229} {\bibfield  {journal}
  {\bibinfo  {journal} {Phys. Rev. B}\ }\textbf {\bibinfo {volume} {43}},\
  \bibinfo {pages} {1229(R)} (\bibinfo {year} {1991})}\BibitemShut {NoStop}%
\bibitem [{\citenamefont {Shinomori}\ \emph {et~al.}(2002)\citenamefont
  {Shinomori}, \citenamefont {Okimoto}, \citenamefont {Kawasaki},\ and\
  \citenamefont {Tokura}}]{shi02}%
  \BibitemOpen
  \bibfield  {author} {\bibinfo {author} {\bibfnamefont {S.}~\bibnamefont
  {Shinomori}}, \bibinfo {author} {\bibfnamefont {Y.}~\bibnamefont {Okimoto}},
  \bibinfo {author} {\bibfnamefont {M.}~\bibnamefont {Kawasaki}},\ and\
  \bibinfo {author} {\bibfnamefont {Y.}~\bibnamefont {Tokura}},\ }\bibfield
  {title} {\bibinfo {title} {Insulator–metal transition in
  la$_{2-x}$sr$_x$nio$_4$},\ }\href {https://doi.org/10.1143/JPSJ.71.705}
  {\bibfield  {journal} {\bibinfo  {journal} {J. Phys. Soc. Jpn.}\ }\textbf
  {\bibinfo {volume} {71}},\ \bibinfo {pages} {705} (\bibinfo {year}
  {2002})}\BibitemShut {NoStop}%
\bibitem [{\citenamefont {Bansal}\ \emph {et~al.}(2023)\citenamefont {Bansal},
  \citenamefont {Maurya}, \citenamefont {Ali}, \citenamefont {Reddy},\ and\
  \citenamefont {Singh}}]{ban23}%
  \BibitemOpen
  \bibfield  {author} {\bibinfo {author} {\bibfnamefont {S.}~\bibnamefont
  {Bansal}}, \bibinfo {author} {\bibfnamefont {R.~K.}\ \bibnamefont {Maurya}},
  \bibinfo {author} {\bibfnamefont {A.}~\bibnamefont {Ali}}, \bibinfo {author}
  {\bibfnamefont {B.~H.}\ \bibnamefont {Reddy}},\ and\ \bibinfo {author}
  {\bibfnamefont {R.~S.}\ \bibnamefont {Singh}},\ }\bibfield  {title} {\bibinfo
  {title} {Role of electron correlation and disorder on the electronic
  structure of layered nickelate
  ${({\mathrm{La}}_{0.5}{\mathrm{Sr}}_{0.5})}_{2}{\mathrm{nio}}_{4}$},\ }\href
  {https://doi.org/10.1103/PhysRevMaterials.7.064007} {\bibfield  {journal}
  {\bibinfo  {journal} {Phys. Rev. Mater.}\ }\textbf {\bibinfo {volume} {7}},\
  \bibinfo {pages} {064007} (\bibinfo {year} {2023})}\BibitemShut {NoStop}%
\bibitem [{\citenamefont {Wissel}\ \emph {et~al.}(2020)\citenamefont {Wissel},
  \citenamefont {Malik}, \citenamefont {Vasala}, \citenamefont {Plana-Ruiz},
  \citenamefont {Kolb}, \citenamefont {Slater}, \citenamefont {da~Silva},
  \citenamefont {Alff}, \citenamefont {Rohrer},\ and\ \citenamefont
  {Clemens}}]{wis20}%
  \BibitemOpen
  \bibfield  {author} {\bibinfo {author} {\bibfnamefont {K.}~\bibnamefont
  {Wissel}}, \bibinfo {author} {\bibfnamefont {A.~M.}\ \bibnamefont {Malik}},
  \bibinfo {author} {\bibfnamefont {S.}~\bibnamefont {Vasala}}, \bibinfo
  {author} {\bibfnamefont {S.}~\bibnamefont {Plana-Ruiz}}, \bibinfo {author}
  {\bibfnamefont {U.}~\bibnamefont {Kolb}}, \bibinfo {author} {\bibfnamefont
  {P.~R.}\ \bibnamefont {Slater}}, \bibinfo {author} {\bibfnamefont
  {I.}~\bibnamefont {da~Silva}}, \bibinfo {author} {\bibfnamefont
  {L.}~\bibnamefont {Alff}}, \bibinfo {author} {\bibfnamefont {J.}~\bibnamefont
  {Rohrer}},\ and\ \bibinfo {author} {\bibfnamefont {O.}~\bibnamefont
  {Clemens}},\ }\bibfield  {title} {\bibinfo {title} {Topochemical reduction of
  la$_2$nio$_3$f$_2$: The first ni-based ruddlesden–popper n = 1 t'-type
  structure and the impact of reduction on magnetic ordering},\ }\href
  {https://doi.org/10.1021/acs.chemmater.0c00193} {\bibfield  {journal}
  {\bibinfo  {journal} {Chem. Mater.}\ }\textbf {\bibinfo {volume} {32}},\
  \bibinfo {pages} {3160} (\bibinfo {year} {2020})}\BibitemShut {NoStop}%
\bibitem [{\citenamefont {Bernardini}\ \emph {et~al.}(2021)\citenamefont
  {Bernardini}, \citenamefont {Demourgues},\ and\ \citenamefont
  {Cano}}]{ber21}%
  \BibitemOpen
  \bibfield  {author} {\bibinfo {author} {\bibfnamefont {F.}~\bibnamefont
  {Bernardini}}, \bibinfo {author} {\bibfnamefont {A.}~\bibnamefont
  {Demourgues}},\ and\ \bibinfo {author} {\bibfnamefont {A.}~\bibnamefont
  {Cano}},\ }\bibfield  {title} {\bibinfo {title} {Single-layer t'-type
  nickelates: ${\mathrm{ni}}^{1+}$ is ${\mathrm{ni}}^{1+}$},\ }\href
  {https://doi.org/10.1103/PhysRevMaterials.5.L061801} {\bibfield  {journal}
  {\bibinfo  {journal} {Phys. Rev. Mater.}\ }\textbf {\bibinfo {volume} {5}},\
  \bibinfo {pages} {L061801} (\bibinfo {year} {2021})}\BibitemShut {NoStop}%
\bibitem [{\citenamefont {Lechermann}\ \emph {et~al.}(2019)\citenamefont
  {Lechermann}, \citenamefont {K\"orner}, \citenamefont {Urban},\ and\
  \citenamefont {Els\"asser}}]{lec19}%
  \BibitemOpen
  \bibfield  {author} {\bibinfo {author} {\bibfnamefont {F.}~\bibnamefont
  {Lechermann}}, \bibinfo {author} {\bibfnamefont {W.}~\bibnamefont
  {K\"orner}}, \bibinfo {author} {\bibfnamefont {D.~F.}\ \bibnamefont
  {Urban}},\ and\ \bibinfo {author} {\bibfnamefont {C.}~\bibnamefont
  {Els\"asser}},\ }\bibfield  {title} {\bibinfo {title} {Interplay of
  charge-transfer and mott-hubbard physics approached by an efficient
  combination of self-interaction correction and dynamical mean-field theory},\
  }\href {https://doi.org/10.1103/PhysRevB.100.115125} {\bibfield  {journal}
  {\bibinfo  {journal} {Phys. Rev. B}\ }\textbf {\bibinfo {volume} {100}},\
  \bibinfo {pages} {115125} (\bibinfo {year} {2019})}\BibitemShut {NoStop}%
\bibitem [{\citenamefont {Elsasser}\ \emph {et~al.}(1990)\citenamefont
  {Elsasser}, \citenamefont {Takeuchi}, \citenamefont {Ho}, \citenamefont
  {Chan}, \citenamefont {Braun},\ and\ \citenamefont {Fahnle}}]{elsaesser90}%
  \BibitemOpen
  \bibfield  {author} {\bibinfo {author} {\bibfnamefont {C.}~\bibnamefont
  {Elsasser}}, \bibinfo {author} {\bibfnamefont {N.}~\bibnamefont {Takeuchi}},
  \bibinfo {author} {\bibfnamefont {K.~M.}\ \bibnamefont {Ho}}, \bibinfo
  {author} {\bibfnamefont {C.~T.}\ \bibnamefont {Chan}}, \bibinfo {author}
  {\bibfnamefont {P.}~\bibnamefont {Braun}},\ and\ \bibinfo {author}
  {\bibfnamefont {M.}~\bibnamefont {Fahnle}},\ }\bibfield  {title} {\bibinfo
  {title} {Relativistic effects on ground state properties of 4d and 5d
  transition metals},\ }\href {https://doi.org/10.1088/0953-8984/2/19/006}
  {\bibfield  {journal} {\bibinfo  {journal} {Journal of Physics: Condensed
  Matter}\ }\textbf {\bibinfo {volume} {2}},\ \bibinfo {pages} {4371} (\bibinfo
  {year} {1990})}\BibitemShut {NoStop}%
\bibitem [{\citenamefont {Lechermann}\ \emph {et~al.}(2002)\citenamefont
  {Lechermann}, \citenamefont {Welsch}, \citenamefont {Els\"asser},
  \citenamefont {Ederer}, \citenamefont {F\"ahnle}, \citenamefont {Sanchez},\
  and\ \citenamefont {Meyer}}]{lechermann02}%
  \BibitemOpen
  \bibfield  {author} {\bibinfo {author} {\bibfnamefont {F.}~\bibnamefont
  {Lechermann}}, \bibinfo {author} {\bibfnamefont {F.}~\bibnamefont {Welsch}},
  \bibinfo {author} {\bibfnamefont {C.}~\bibnamefont {Els\"asser}}, \bibinfo
  {author} {\bibfnamefont {C.}~\bibnamefont {Ederer}}, \bibinfo {author}
  {\bibfnamefont {M.}~\bibnamefont {F\"ahnle}}, \bibinfo {author}
  {\bibfnamefont {J.~M.}\ \bibnamefont {Sanchez}},\ and\ \bibinfo {author}
  {\bibfnamefont {B.}~\bibnamefont {Meyer}},\ }\bibfield  {title} {\bibinfo
  {title} {Density-functional study of ${\mathrm{fe}}_{3}\mathrm{Al}:$ lsda
  versus gga},\ }\href {https://doi.org/10.1103/PhysRevB.65.132104} {\bibfield
  {journal} {\bibinfo  {journal} {Phys. Rev. B}\ }\textbf {\bibinfo {volume}
  {65}},\ \bibinfo {pages} {132104} (\bibinfo {year} {2002})}\BibitemShut
  {NoStop}%
\bibitem [{\citenamefont {Meyer}\ \emph {et~al.}(1998)\citenamefont {Meyer},
  \citenamefont {Els\"{a}sser}, \citenamefont {Lechermann},\ and\ \citenamefont
  {F\"{a}hnle}}]{mbpp_code}%
  \BibitemOpen
  \bibfield  {author} {\bibinfo {author} {\bibfnamefont {B.}~\bibnamefont
  {Meyer}}, \bibinfo {author} {\bibfnamefont {C.}~\bibnamefont {Els\"{a}sser}},
  \bibinfo {author} {\bibfnamefont {F.}~\bibnamefont {Lechermann}},\ and\
  \bibinfo {author} {\bibfnamefont {M.}~\bibnamefont {F\"{a}hnle}},\
  }\href@noop {} {\emph {\bibinfo {title} {FORTRAN 90 Program for
  Mixed-Basis-Pseudopotential Calculations for Crystals}}},\ \bibinfo
  {organization} {Max-Planck-Institut f\"{u}r Metallforschung, Stuttgart}
  (\bibinfo {year} {1998})\BibitemShut {NoStop}%
\bibitem [{\citenamefont {K\"orner}\ and\ \citenamefont
  {Els\"asser}(2010)}]{korner10}%
  \BibitemOpen
  \bibfield  {author} {\bibinfo {author} {\bibfnamefont {W.}~\bibnamefont
  {K\"orner}}\ and\ \bibinfo {author} {\bibfnamefont {C.}~\bibnamefont
  {Els\"asser}},\ }\bibfield  {title} {\bibinfo {title} {First-principles
  density functional study of dopant elements at grain boundaries in zno},\
  }\href {https://doi.org/10.1103/PhysRevB.81.085324} {\bibfield  {journal}
  {\bibinfo  {journal} {Phys. Rev. B}\ }\textbf {\bibinfo {volume} {81}},\
  \bibinfo {pages} {085324} (\bibinfo {year} {2010})}\BibitemShut {NoStop}%
\bibitem [{\citenamefont {Werner}\ \emph {et~al.}(2006)\citenamefont {Werner},
  \citenamefont {Comanac}, \citenamefont {de' Medici}, \citenamefont {Troyer},\
  and\ \citenamefont {Millis}}]{werner06}%
  \BibitemOpen
  \bibfield  {author} {\bibinfo {author} {\bibfnamefont {P.}~\bibnamefont
  {Werner}}, \bibinfo {author} {\bibfnamefont {A.}~\bibnamefont {Comanac}},
  \bibinfo {author} {\bibfnamefont {L.}~\bibnamefont {de' Medici}}, \bibinfo
  {author} {\bibfnamefont {M.}~\bibnamefont {Troyer}},\ and\ \bibinfo {author}
  {\bibfnamefont {A.~J.}\ \bibnamefont {Millis}},\ }\bibfield  {title}
  {\bibinfo {title} {Continuous-time solver for quantum impurity models},\
  }\href {https://doi.org/10.1103/PhysRevLett.97.076405} {\bibfield  {journal}
  {\bibinfo  {journal} {Phys. Rev. Lett.}\ }\textbf {\bibinfo {volume} {97}},\
  \bibinfo {pages} {076405} (\bibinfo {year} {2006})}\BibitemShut {NoStop}%
\bibitem [{\citenamefont {Parcollet}\ \emph {et~al.}(2015)\citenamefont
  {Parcollet}, \citenamefont {Ferrero}, \citenamefont {Ayral}, \citenamefont
  {Hafermann}, \citenamefont {Krivenko}, \citenamefont {Messio},\ and\
  \citenamefont {Seth}}]{parcollet15}%
  \BibitemOpen
  \bibfield  {author} {\bibinfo {author} {\bibfnamefont {O.}~\bibnamefont
  {Parcollet}}, \bibinfo {author} {\bibfnamefont {M.}~\bibnamefont {Ferrero}},
  \bibinfo {author} {\bibfnamefont {T.}~\bibnamefont {Ayral}}, \bibinfo
  {author} {\bibfnamefont {H.}~\bibnamefont {Hafermann}}, \bibinfo {author}
  {\bibfnamefont {I.}~\bibnamefont {Krivenko}}, \bibinfo {author}
  {\bibfnamefont {L.}~\bibnamefont {Messio}},\ and\ \bibinfo {author}
  {\bibfnamefont {P.}~\bibnamefont {Seth}},\ }\bibfield  {title} {\bibinfo
  {title} {Triqs: A toolbox for research on interacting quantum systems},\
  }\href {https://doi.org/10.1016/j.cpc.2015.04.023} {\bibfield  {journal}
  {\bibinfo  {journal} {Computer Physics Communications}\ }\textbf {\bibinfo
  {volume} {196}},\ \bibinfo {pages} {398} (\bibinfo {year}
  {2015})}\BibitemShut {NoStop}%
\bibitem [{\citenamefont {Seth}\ \emph {et~al.}(2016)\citenamefont {Seth},
  \citenamefont {Krivenko}, \citenamefont {Ferrero},\ and\ \citenamefont
  {Parcollet}}]{seth16}%
  \BibitemOpen
  \bibfield  {author} {\bibinfo {author} {\bibfnamefont {P.}~\bibnamefont
  {Seth}}, \bibinfo {author} {\bibfnamefont {I.}~\bibnamefont {Krivenko}},
  \bibinfo {author} {\bibfnamefont {M.}~\bibnamefont {Ferrero}},\ and\ \bibinfo
  {author} {\bibfnamefont {O.}~\bibnamefont {Parcollet}},\ }\bibfield  {title}
  {\bibinfo {title} {Triqs/cthyb: A continuous-time quantum monte carlo
  hybridisation expansion solver for quantum impurity problems},\ }\href
  {https://doi.org/10.1016/j.cpc.2015.10.023} {\bibfield  {journal} {\bibinfo
  {journal} {Comput. Phys. Commun.}\ }\textbf {\bibinfo {volume} {200}},\
  \bibinfo {pages} {274} (\bibinfo {year} {2016})}\BibitemShut {NoStop}%
\bibitem [{\citenamefont {Amadon}\ \emph {et~al.}(2008)\citenamefont {Amadon},
  \citenamefont {Lechermann}, \citenamefont {Georges}, \citenamefont {Jollet},
  \citenamefont {Wehling},\ and\ \citenamefont {Lichtenstein}}]{amadon08}%
  \BibitemOpen
  \bibfield  {author} {\bibinfo {author} {\bibfnamefont {B.}~\bibnamefont
  {Amadon}}, \bibinfo {author} {\bibfnamefont {F.}~\bibnamefont {Lechermann}},
  \bibinfo {author} {\bibfnamefont {A.}~\bibnamefont {Georges}}, \bibinfo
  {author} {\bibfnamefont {F.}~\bibnamefont {Jollet}}, \bibinfo {author}
  {\bibfnamefont {T.~O.}\ \bibnamefont {Wehling}},\ and\ \bibinfo {author}
  {\bibfnamefont {A.~I.}\ \bibnamefont {Lichtenstein}},\ }\bibfield  {title}
  {\bibinfo {title} {Plane-wave based electronic structure calculations for
  correlated materials using dynamical mean-field theory and projected local
  orbitals},\ }\href {https://doi.org/10.1103/PhysRevB.77.205112} {\bibfield
  {journal} {\bibinfo  {journal} {Phys. Rev. B}\ }\textbf {\bibinfo {volume}
  {77}},\ \bibinfo {pages} {205112} (\bibinfo {year} {2008})}\BibitemShut
  {NoStop}%
\bibitem [{\citenamefont {Anisimov}\ \emph {et~al.}(1993)\citenamefont
  {Anisimov}, \citenamefont {Solovyev}, \citenamefont {Korotin}, \citenamefont
  {Czy\ifmmode~\dot{z}\else \.{z}\fi{}yk},\ and\ \citenamefont
  {Sawatzky}}]{anisimov93}%
  \BibitemOpen
  \bibfield  {author} {\bibinfo {author} {\bibfnamefont {V.~I.}\ \bibnamefont
  {Anisimov}}, \bibinfo {author} {\bibfnamefont {I.~V.}\ \bibnamefont
  {Solovyev}}, \bibinfo {author} {\bibfnamefont {M.~A.}\ \bibnamefont
  {Korotin}}, \bibinfo {author} {\bibfnamefont {M.~T.}\ \bibnamefont
  {Czy\ifmmode~\dot{z}\else \.{z}\fi{}yk}},\ and\ \bibinfo {author}
  {\bibfnamefont {G.~A.}\ \bibnamefont {Sawatzky}},\ }\bibfield  {title}
  {\bibinfo {title} {Density-functional theory and nio photoemission spectra},\
  }\href {https://doi.org/10.1103/PhysRevB.48.16929} {\bibfield  {journal}
  {\bibinfo  {journal} {Phys. Rev. B}\ }\textbf {\bibinfo {volume} {48}},\
  \bibinfo {pages} {16929} (\bibinfo {year} {1993})}\BibitemShut {NoStop}%
\bibitem [{\citenamefont {Zaanen}\ \emph {et~al.}(1985)\citenamefont {Zaanen},
  \citenamefont {Sawatzky},\ and\ \citenamefont {Allen}}]{zaa85}%
  \BibitemOpen
  \bibfield  {author} {\bibinfo {author} {\bibfnamefont {J.}~\bibnamefont
  {Zaanen}}, \bibinfo {author} {\bibfnamefont {G.~A.}\ \bibnamefont
  {Sawatzky}},\ and\ \bibinfo {author} {\bibfnamefont {J.~W.}\ \bibnamefont
  {Allen}},\ }\bibfield  {title} {\bibinfo {title} {Band gaps and electronic
  structure of transition-metal compounds},\ }\href
  {https://doi.org/10.1103/PhysRevLett.55.418} {\bibfield  {journal} {\bibinfo
  {journal} {Phys. Rev. Lett.}\ }\textbf {\bibinfo {volume} {55}},\ \bibinfo
  {pages} {418} (\bibinfo {year} {1985})}\BibitemShut {NoStop}%
\bibitem [{\citenamefont {Lechermann}\ \emph {et~al.}(2023)\citenamefont
  {Lechermann}, \citenamefont {Gondolf}, \citenamefont {B\"otzel},\ and\
  \citenamefont {Eremin}}]{lec23}%
  \BibitemOpen
  \bibfield  {author} {\bibinfo {author} {\bibfnamefont {F.}~\bibnamefont
  {Lechermann}}, \bibinfo {author} {\bibfnamefont {J.}~\bibnamefont {Gondolf}},
  \bibinfo {author} {\bibfnamefont {S.}~\bibnamefont {B\"otzel}},\ and\
  \bibinfo {author} {\bibfnamefont {I.~M.}\ \bibnamefont {Eremin}},\ }\bibfield
   {title} {\bibinfo {title} {Electronic correlations and superconducting
  instability in ${\mathrm{la}}_{3}{\mathrm{ni}}_{2}{\mathrm{o}}_{7}$ under
  high pressure},\ }\href {https://doi.org/10.1103/PhysRevB.108.L201121}
  {\bibfield  {journal} {\bibinfo  {journal} {Phys. Rev. B}\ }\textbf {\bibinfo
  {volume} {108}},\ \bibinfo {pages} {L201121} (\bibinfo {year}
  {2023})}\BibitemShut {NoStop}%
\bibitem [{\citenamefont {Lechermann}\ \emph {et~al.}(2025)\citenamefont
  {Lechermann}, \citenamefont {B\"otzel},\ and\ \citenamefont
  {Eremin}}]{lec25}%
  \BibitemOpen
  \bibfield  {author} {\bibinfo {author} {\bibfnamefont {F.}~\bibnamefont
  {Lechermann}}, \bibinfo {author} {\bibfnamefont {S.}~\bibnamefont
  {B\"otzel}},\ and\ \bibinfo {author} {\bibfnamefont {I.~M.}\ \bibnamefont
  {Eremin}},\ }\bibfield  {title} {\bibinfo {title} {Interplay of
  orbital-selective mott criticality and flat-band physics in
  ${\mathrm{la}}_{3}{\mathrm{ni}}_{2}{\mathrm{o}}_{6}$},\ }\href
  {https://doi.org/10.1103/3r7g-89r8} {\bibfield  {journal} {\bibinfo
  {journal} {Phys. Rev. B}\ }\textbf {\bibinfo {volume} {112}},\ \bibinfo
  {pages} {245125} (\bibinfo {year} {2025})}\BibitemShut {NoStop}%
\bibitem [{\citenamefont {Hayward}\ \emph {et~al.}(1999)\citenamefont
  {Hayward}, \citenamefont {Green}, \citenamefont {Rosseinsky},\ and\
  \citenamefont {Sloan}}]{hay99}%
  \BibitemOpen
  \bibfield  {author} {\bibinfo {author} {\bibfnamefont {M.~A.}\ \bibnamefont
  {Hayward}}, \bibinfo {author} {\bibfnamefont {M.~A.}\ \bibnamefont {Green}},
  \bibinfo {author} {\bibfnamefont {M.~J.}\ \bibnamefont {Rosseinsky}},\ and\
  \bibinfo {author} {\bibfnamefont {J.}~\bibnamefont {Sloan}},\ }\bibfield
  {title} {\bibinfo {title} {Sodium hydride as a powerful reducing agent for
  topotactic oxide deintercalation: Synthesis and characterization of the
  nickel(i) oxide lanio$_2$},\ }\href {https://doi.org/10.1021/ja991573i}
  {\bibfield  {journal} {\bibinfo  {journal} {J. Am. Chem. Soc.}\ }\textbf
  {\bibinfo {volume} {121}},\ \bibinfo {pages} {8843} (\bibinfo {year}
  {1999})}\BibitemShut {NoStop}%
\bibitem [{\citenamefont {apRoberts Warren}\ \emph {et~al.}(2013)\citenamefont
  {apRoberts Warren}, \citenamefont {Crocker}, \citenamefont {Dioguardi},
  \citenamefont {Shirer}, \citenamefont {Poltavets}, \citenamefont
  {Greenblatt}, \citenamefont {Klavins},\ and\ \citenamefont {Curro}}]{apR13}%
  \BibitemOpen
  \bibfield  {author} {\bibinfo {author} {\bibfnamefont {N.}~\bibnamefont
  {apRoberts Warren}}, \bibinfo {author} {\bibfnamefont {J.}~\bibnamefont
  {Crocker}}, \bibinfo {author} {\bibfnamefont {A.~P.}\ \bibnamefont
  {Dioguardi}}, \bibinfo {author} {\bibfnamefont {K.~R.}\ \bibnamefont
  {Shirer}}, \bibinfo {author} {\bibfnamefont {V.~V.}\ \bibnamefont
  {Poltavets}}, \bibinfo {author} {\bibfnamefont {M.}~\bibnamefont
  {Greenblatt}}, \bibinfo {author} {\bibfnamefont {P.}~\bibnamefont
  {Klavins}},\ and\ \bibinfo {author} {\bibfnamefont {N.~J.}\ \bibnamefont
  {Curro}},\ }\bibfield  {title} {\bibinfo {title} {Nmr evidence for spin
  fluctuations in the bilayer nickelate la${}_{3}$ni${}_{2}$o${}_{6}$},\ }\href
  {https://doi.org/10.1103/PhysRevB.88.075124} {\bibfield  {journal} {\bibinfo
  {journal} {Phys. Rev. B}\ }\textbf {\bibinfo {volume} {88}},\ \bibinfo
  {pages} {075124} (\bibinfo {year} {2013})}\BibitemShut {NoStop}%
\bibitem [{\citenamefont {Zhang}\ \emph {et~al.}(2017)\citenamefont {Zhang},
  \citenamefont {Botana}, \citenamefont {Freeland}, \citenamefont {Phelan},
  \citenamefont {Zheng}, \citenamefont {Pardo}, \citenamefont {Norma},\ and\
  \citenamefont {Mitchell}}]{zhang17}%
  \BibitemOpen
  \bibfield  {author} {\bibinfo {author} {\bibfnamefont {J.}~\bibnamefont
  {Zhang}}, \bibinfo {author} {\bibfnamefont {A.~S.}\ \bibnamefont {Botana}},
  \bibinfo {author} {\bibfnamefont {J.~W.}\ \bibnamefont {Freeland}}, \bibinfo
  {author} {\bibfnamefont {D.}~\bibnamefont {Phelan}}, \bibinfo {author}
  {\bibfnamefont {H.}~\bibnamefont {Zheng}}, \bibinfo {author} {\bibfnamefont
  {V.}~\bibnamefont {Pardo}}, \bibinfo {author} {\bibfnamefont {M.~R.}\
  \bibnamefont {Norma}},\ and\ \bibinfo {author} {\bibfnamefont {J.~F.}\
  \bibnamefont {Mitchell}},\ }\bibfield  {title} {\bibinfo {title} {Large
  orbital polarization in a metallic square-planar nickelate},\ }\href
  {https://doi.org/10.1038/nphys4149} {\bibfield  {journal} {\bibinfo
  {journal} {Nature Phys.}\ }\textbf {\bibinfo {volume} {13}},\ \bibinfo
  {pages} {864} (\bibinfo {year} {2017})}\BibitemShut {NoStop}%
\bibitem [{\citenamefont {Pan}\ \emph {et~al.}(2021)\citenamefont {Pan},
  \citenamefont {Segedin}, \citenamefont {LaBollita}, \citenamefont {Song},
  \citenamefont {Nica}, \citenamefont {Goodge}, \citenamefont {Pierce},
  \citenamefont {Doyle}, \citenamefont {Novakov}, \citenamefont {Carrizales},
  \citenamefont {N'Diaye}, \citenamefont {Shafer}, \citenamefont {Paik},
  \citenamefont {Heron}, \citenamefont {Mason}, \citenamefont {Yacoby},
  \citenamefont {Kourkoutis}, \citenamefont {Erten}, \citenamefont {Brooks},
  \citenamefont {Botana},\ and\ \citenamefont {Mundy}}]{pan21}%
  \BibitemOpen
  \bibfield  {author} {\bibinfo {author} {\bibfnamefont {G.~A.}\ \bibnamefont
  {Pan}}, \bibinfo {author} {\bibfnamefont {D.~F.}\ \bibnamefont {Segedin}},
  \bibinfo {author} {\bibfnamefont {H.}~\bibnamefont {LaBollita}}, \bibinfo
  {author} {\bibfnamefont {Q.}~\bibnamefont {Song}}, \bibinfo {author}
  {\bibfnamefont {E.~M.}\ \bibnamefont {Nica}}, \bibinfo {author}
  {\bibfnamefont {B.~H.}\ \bibnamefont {Goodge}}, \bibinfo {author}
  {\bibfnamefont {A.~T.}\ \bibnamefont {Pierce}}, \bibinfo {author}
  {\bibfnamefont {S.}~\bibnamefont {Doyle}}, \bibinfo {author} {\bibfnamefont
  {S.}~\bibnamefont {Novakov}}, \bibinfo {author} {\bibfnamefont {D.~C.}\
  \bibnamefont {Carrizales}}, \bibinfo {author} {\bibfnamefont {A.~T.}\
  \bibnamefont {N'Diaye}}, \bibinfo {author} {\bibfnamefont {P.}~\bibnamefont
  {Shafer}}, \bibinfo {author} {\bibfnamefont {H.}~\bibnamefont {Paik}},
  \bibinfo {author} {\bibfnamefont {J.~T.}\ \bibnamefont {Heron}}, \bibinfo
  {author} {\bibfnamefont {J.~A.}\ \bibnamefont {Mason}}, \bibinfo {author}
  {\bibfnamefont {A.}~\bibnamefont {Yacoby}}, \bibinfo {author} {\bibfnamefont
  {L.~F.}\ \bibnamefont {Kourkoutis}}, \bibinfo {author} {\bibfnamefont
  {O.}~\bibnamefont {Erten}}, \bibinfo {author} {\bibfnamefont {C.~M.}\
  \bibnamefont {Brooks}}, \bibinfo {author} {\bibfnamefont {A.~S.}\
  \bibnamefont {Botana}},\ and\ \bibinfo {author} {\bibfnamefont {J.~A.}\
  \bibnamefont {Mundy}},\ }\bibfield  {title} {\bibinfo {title}
  {Superconductivity in a quintuple-layer square-planar nickelate},\ }\href
  {https://doi.org/10.1038/s41563-021-01142-9} {\bibfield  {journal} {\bibinfo
  {journal} {Nature Materials}\ }\textbf {\bibinfo {volume} {21}},\ \bibinfo
  {pages} {160} (\bibinfo {year} {2021})}\BibitemShut {NoStop}%
\bibitem [{\citenamefont {Lechermann}(2021)}]{lec21}%
  \BibitemOpen
  \bibfield  {author} {\bibinfo {author} {\bibfnamefont {F.}~\bibnamefont
  {Lechermann}},\ }\bibfield  {title} {\bibinfo {title} {Doping-dependent
  character and possible magnetic ordering of ${\mathrm{ndnio}}_{2}$},\ }\href
  {https://doi.org/10.1103/PhysRevMaterials.5.044803} {\bibfield  {journal}
  {\bibinfo  {journal} {Phys. Rev. Mater.}\ }\textbf {\bibinfo {volume} {5}},\
  \bibinfo {pages} {044803} (\bibinfo {year} {2021})}\BibitemShut {NoStop}%
\bibitem [{\citenamefont {Andersen}\ \emph {et~al.}(2007)\citenamefont
  {Andersen}, \citenamefont {Hirschfeld}, \citenamefont {Kampf},\ and\
  \citenamefont {Schmid}}]{ande07}%
  \BibitemOpen
  \bibfield  {author} {\bibinfo {author} {\bibfnamefont {B.~M.}\ \bibnamefont
  {Andersen}}, \bibinfo {author} {\bibfnamefont {P.~J.}\ \bibnamefont
  {Hirschfeld}}, \bibinfo {author} {\bibfnamefont {A.~P.}\ \bibnamefont
  {Kampf}},\ and\ \bibinfo {author} {\bibfnamefont {M.}~\bibnamefont
  {Schmid}},\ }\bibfield  {title} {\bibinfo {title} {Disorder-induced static
  antiferromagnetism in cuprate superconductors},\ }\href
  {https://doi.org/10.1103/PhysRevLett.99.147002} {\bibfield  {journal}
  {\bibinfo  {journal} {Phys. Rev. Lett.}\ }\textbf {\bibinfo {volume} {99}},\
  \bibinfo {pages} {147002} (\bibinfo {year} {2007})}\BibitemShut {NoStop}%
\bibitem [{\citenamefont {Behrmann}\ and\ \citenamefont
  {Lechermann}(2015)}]{beh15}%
  \BibitemOpen
  \bibfield  {author} {\bibinfo {author} {\bibfnamefont {M.}~\bibnamefont
  {Behrmann}}\ and\ \bibinfo {author} {\bibfnamefont {F.}~\bibnamefont
  {Lechermann}},\ }\bibfield  {title} {\bibinfo {title} {Interface exchange
  processes in ${\mathrm{laalo}}_{3}/\mathrm{Sr}{\mathrm{tio}}_{3}$ induced by
  oxygen vacancies},\ }\href {https://doi.org/10.1103/PhysRevB.92.125148}
  {\bibfield  {journal} {\bibinfo  {journal} {Phys. Rev. B}\ }\textbf {\bibinfo
  {volume} {92}},\ \bibinfo {pages} {125148} (\bibinfo {year}
  {2015})}\BibitemShut {NoStop}%
\end{thebibliography}%
\end{document}